# Structural changes in $Ge_{1-x}Sn_x$ and $Si_{1-x-y}Ge_ySn_x$ thin films on SOI substrates treated by pulse laser annealing

O. Steuer[1,4], D. Schwarz[2], M. Oehme[2], F. Bärwolf[3], Y. Cheng[1], F. Ganss[1], R. Hübner[1], R. Heller[1], S. Zhou[1], M. Helm[1,6], G. Cuniberti[4], Y. M. Georgiev[1,5], and S. Prucnal[1]

[1] Institute of Ion Beam Physics and Materials Research, Helmholtz-Zentrum Dresden-Rossendorf, Bautzner Landstrasse 400, 01328 Dresden, Germany

[2] University of Stuttgart, Institute of Semiconductor Engineering, Pfaffenwaldring 47, 70569 Stuttgart, Germany

[3] IHP - Leibniz-Institut für innovative Mikroelektronik, Im Technologiepark 25, 15236 Frankfurt (Oder), Germany

[4] Institute of Materials Science, Technische Universität Dresden, Budapester Str. 27, 01069 Dresden, Germany

[5] Institute of Electronics, Bulgarian Academy of Sciences, 72, Tsarigradsko Chausse Blvd.,1784 Sofia, Bulgaria

[6] Center for Advancing Electronics Dresden, Technische Universität Dresden, Helmholtzstraße 18, 01062 Dresden, Germany

E-mail: o.steuer@hzdr.de, s.prucnal@hzdr.de

## Abstract

$Ge_{1-x}Sn_x$ and $Si_{1-x-y}Ge_ySn_x$ alloys are promising materials for future opto- and nanoelectronics applications. These alloys enable effective band-gap engineering, broad adjustability of their lattice parameter, exhibit much higher carrier mobility than pure Si, and are compatible with the CMOS technology. Unfortunately, the equilibrium solid solubility of Sn in $Si_{1-x}Ge_x$ is less than 1% and the pseudomorphic growth of $Si_{1-x-y}Ge_ySn_x$ on Ge or Si can cause in-plane compressive strain in the grown layer, degrading the superior properties of these alloys. Therefore, post-growth strain engineering by ultrafast non-equilibrium thermal treatments like pulse laser annealing (PLA) is needed to improve the layer quality. In this article, $Ge_{0.94}Sn_{0.06}$ and $Si_{0.14}Ge_{0.8}Sn_{0.06}$ thin films grown on silicon-on-insulator substrates by molecular beam epitaxy were post-growth thermally treated by PLA. The material is analyzed before and after the thermal treatments by transmission electron microscopy, X-ray diffraction (XRD), Rutherford backscattering spectrometry, secondary ion mass spectrometry, and Hall-effect measurements. It is shown that after annealing, the material is single-crystalline with improved crystallinity than the as-grown layer. This is reflected in a significantly increased XRD reflection intensity, well-ordered atomic pillars, and increased active carrier concentrations up to $4 \times 10^{19}$ cm$^{-3}$.

## I Introduction

Currently, the industry´s main production lines of integrated circuits (ICs) are predominantly based on group-IV materials such as silicon (Si), germanium (Ge), $Si_{1-x}Ge_x$ or silicon carbide (SiC). Two emerging newcomers in this class are tin- (Sn) containing $Ge_{1-x}Sn_x$ and $Si_{1-x-y}Ge_ySn_x$ alloys. These alloys showed the potential of superior mobilities of up to 6000 cm²/Vs for electrons and 4500 cm²/Vs for holes [1-5]. However, it requires a high material quality to achieve such performances. Typically, both alloys are epitaxially grown on Ge or Ge-buffered substrates by chemical vapor deposition (CVD) [6, 7] or molecular beam epitaxy (MBE) [8, 9]. Ge is used since it is CMOS-compatible, has a relatively large lattice parameter, and allows post-growth thermal treatments under equilibrium conditions [10]. Unfortunately, pure Ge wafers are expensive and the more complex layer stack of Ge-buffered Si substrates complicates the fabrication and characterization of lateral ICs. Furthermore, for many device concepts, insulating substrates are desired [11]. Recently, the first GeSnOI substrates were fabricated by direct wafer bonding and etch-back approach [11-13] or by underetching and layer transfer to insulating substrates [14]. These methods are cost-intensive and technically challenging. A much simpler way to fabricate $Ge_{1-x}Sn_x$ and $Si_{1-x-y}Ge_ySn_x$ on an insulating platform would be the direct growth of these alloys on commercially available ultra-thin silicon-on-insulator (SOI) wafers. In the last years, the feasibility of growing single-crystalline $Ge_{1-x}Sn_x$ directly on Si substrates [15-17] was shown, and defect densities of about ~$10^7$ cm$^{-2}$ were estimated [17]. In this paper, we present the fabrication of $Ge_{0.94}Sn_{0.06}$ and $Si_{0.14}Ge_{0.80}Sn_{0.06}$ on commercially available SOI substrates and apply non-equilibrium post-growth pulse laser annealing (PLA) to mediate between the highly lattice-mismatched alloys and the SOI substrate beneath. Nanosecond PLA helps to improve the crystal structure of the $Ge_{1-x}Sn_x$ and $Si_{1-x-y}Ge_ySn_x$ alloys.

## II Experimental part

A commercially available 300 mm-SOI wafer with a 20 nm-thick top Si and 100 nm-thick $SiO_2$ layer was used as a low-cost substrate. The wafer was diced into 35 x 35 mm² pieces, and the native $SiO_2$ was removed by a 2.5 % HF:DI etching for 15 s. Afterward, four SOI pieces were simultaneously inserted for each of the alloys into a solid-source MBE system with a base pressure of 1 x $10^{-10}$ mbar. The substrate was heated up to 700 °C with a ramp of about 30°K min$^{-1}$. Then, the temperature was kept constant for 5 minutes to eliminate hydrogen dangling bonds on the Si surface created by HF etching [18]. After thermal stabilization, the MBE growth was performed at around 180°C for $Ge_{0.94}Sn_{0.06}$ and 200 °C for $Si_{0.14}Ge_{0.80}Sn_{0.06}$. Both 20 nm-thick layers are *in situ* doped with Sb with a targeted concentration of 5 x $10^{19}$ cm$^{-3}$. The elements Ge, Sn, and Sb were evaporated using Knudsen effusion cells. Si was evaporated via electron beam. Post-growth PLA was performed under atmospheric conditions with single pulses. The PLA tool "Coherent VarioLas ECO 308 5x5 for COMPexPro200" is equipped with a 308 nm wavelength XeCl excimer laser with a fixed pulse length of 28 ns. A "MaxBlack EnergyMax Sensor" measured the laser energy from which the energy density for the homogeneously irradiated 5 x 5 mm² area was calculated.

Cross-sectional high-resolution transmission electron microscopy (HR-TEM) imaging to evaluate the layer crystallinity was performed using an image-$C_s$-corrected Titan 80-300 microscope (FEI) operated at an accelerating voltage of 300 kV. High-angle annular dark-field scanning transmission electron microscopy (HAADF-STEM) imaging and spectrum imaging analysis based on energy-dispersive X-ray spectroscopy (EDXS) were performed at 200 kV with a Talos F200X microscope equipped with a Super-X EDX detector system (FEI). Before (S)TEM analysis, the specimen mounted in a high-visibility low-background holder was placed for 10 s into a Model 1020 Plasma Cleaner (Fischione) to remove potential contamination. Cross-sectional TEM lamellae preparation was done by *in situ* lift-out using a Helios 5 CX focused ion beam (FIB) device (Thermo Fisher, Eindhoven, Netherlands). For each sample, to protect its surface, a carbon cap layer was deposited beginning with electron-beam-assisted and subsequently followed by Ga-FIB-assisted precursor decomposition. Afterward, the TEM lamella was prepared using a 30 keV Ga-FIB with adapted currents. Its transfer to a 3-post copper Omniprobe lift-out grid was done with an EasyLift EX nanomanipulator (Thermo Fisher). To minimize sidewall damage, Ga ions with only 5 keV energy were used for final thinning of the TEM lamella to electron transparency. X-ray diffraction (XRD) was performed on a Rigaku SmartLab X-ray diffractometer system. The system is equipped with a copper (Cu) X-ray tube and a Ge (220) two-bounce monochromator. High-resolution XRD (HR-XRD) $\theta$-2$\theta$ scans were carried out on the symmetrical 0 0 4 reflections. Reciprocal space maps (RSM) were generated for the 0 0 4 and the asymmetrical 2 2 4 reflections by scanning $\omega$ and measuring the diffraction intensity in dependence of 2$\theta$ with the detector in 1D single-exposure mode. The samples were aligned by their 0 0 4 and 2 2 4 Si substrate reflections for these measurements. The lateral and vertical lattice parameters of the epitaxially grown layers were determined from the 2 2 4 RSM. Rutherford backscattering spectrometry (RBS) in random (RBS-R) and channeling (RBS-C) directions were performed using a 2 MV Van-de-Graaff accelerator. The used He$^+$ beam had an energy of 1.7 MeV and beam currents between 10 and 20 nA. An aperture with a diameter of about 1 mm was used. Each measurement was performed with a detector angle of 170°. The obtained RBS spectra were fitted with the software SIMNRA [19]. RBS-C was performed along the [001] crystal axis. Time-of-flight (ToF) secondary ion mass spectrometry (SIMS) measurements were conducted before and after annealing. The used IONTOF V tool is equipped with a ToF mass separator. The material was sputtered by Cs$^+$ with an energy of 500 eV and a sputter area of 400 × 400 µm². The analysis was performed with Bi$_1^+$ having an energy of 15 keV and an area of 200 × 200 µm². Variable-field Hall-effect measurements in van-der-Pauw configuration were performed at room temperature using an HMS 9709A system from LakeShore. Close to the sample corners, 50 nm-thick circularly-shaped Ni contacts with a diameter of 1 mm were fabricated by lithography, 1 % HF:DI oxide etching, thermal evaporation of Ni, and lift-off. The magnetic flux density was varied between -5 and 5 T.

### III. Results and discussion

The microstructure of the as-grown Ge$_{0.94}$Sn$_{0.06}$ and Si$_{0.14}$Ge$_{0.80}$Sn$_{0.06}$ on SOI was investigated by TEM-based analyses, as presented in Fig. 1. The superimposed EDXS-based element distribution maps in Fig. 1 a) and b) confirm the targeted layer stack with a relatively homogeneous distribution of Sn within Ge$_{0.94}$Sn$_{0.06}$ and Si$_{0.14}$Ge$_{0.80}$Sn$_{0.06}$, respectively. As shown in Fig. 1c) and e), the microstructure of Ge$_{0.94}$Sn$_{0.06}$ appears to be partially crystalline, where only some of the defective grains grow up to the sample surface. The other layer regions are characterized by an amorphous microstructure. The presence of such amorphous inclusions indicates an epitaxial breakdown during the growth, which appears when the critical thickness of highly mismatched alloys is exceeded [20]. The Si$_{0.14}$Ge$_{0.80}$Sn$_{0.06}$ layer in Fig. 1 d) and f) is mainly single-crystalline but contains many stacking faults. The top surfaces of Ge$_{0.94}$Sn$_{0.06}$ and Si$_{0.14}$Ge$_{0.80}$Sn$_{0.06}$ have a native oxide layer and seem to be slightly rougher compared to the Ge$_{1-x}$Sn$_x$ and Si$_{1-x-y}$Ge$_y$Sn$_x$/Si interfaces respectively. This suggests a "layer-by-layer" plus "three-dimensional island"-based (Stranski-Krastanov) growth with a small island spacing rather than the targeted "layer-by-layer" (Frank–van der Merwe) growth mechanism [21]. Since the microstructures of the as-grown alloys are defect-rich, post-growth annealing is required to improve the crystallinity. On the other hand, these Sn-containing alloys are metastable and would decompose during long thermal treatments above the growth temperature. Consequently, we decided to use ns-range PLA.

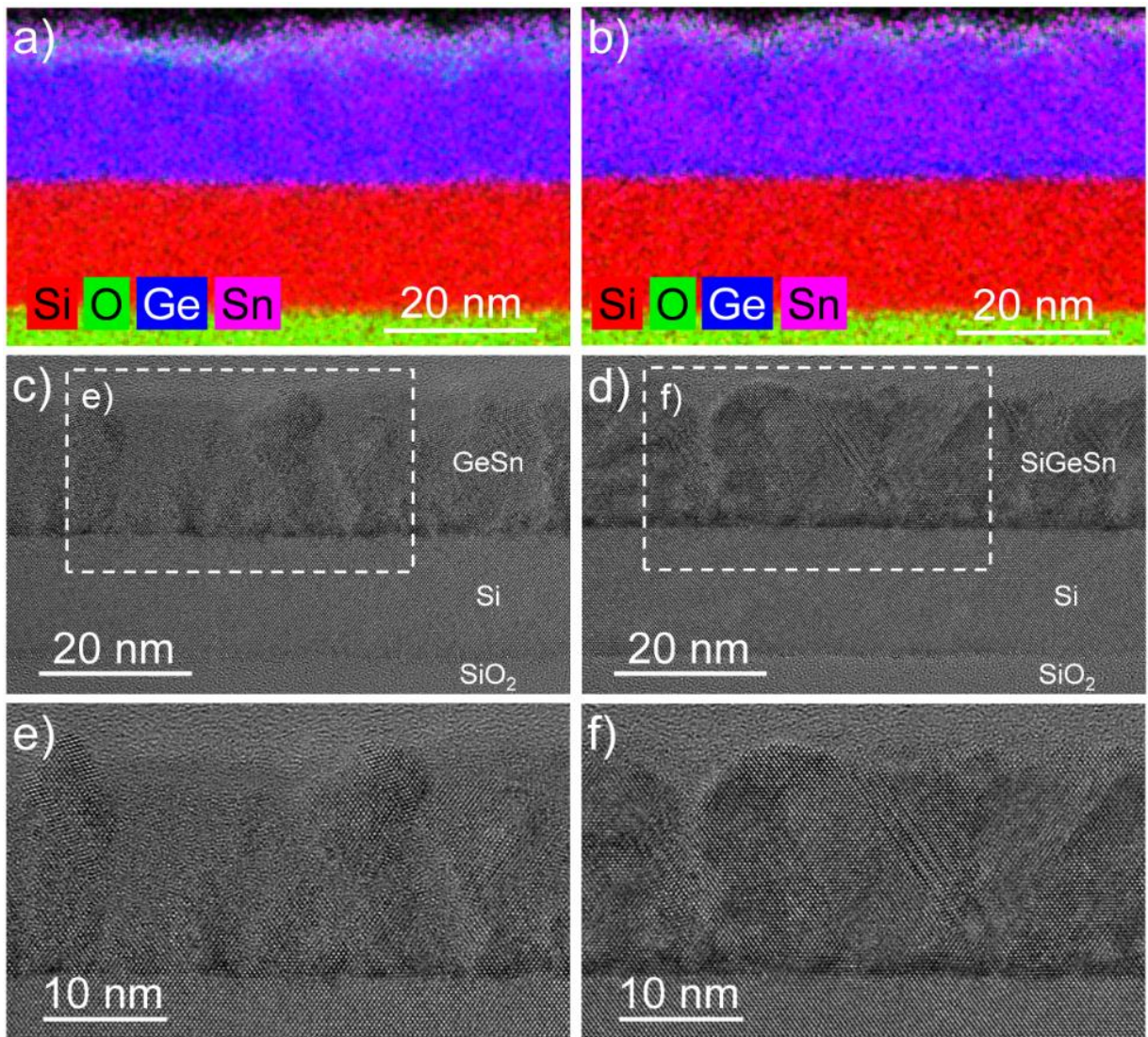

Fig. 1: TEM-based analysis of the as-grown Ge$_{0.94}$Sn$_{0.06}$ on SOI in a), c), and e) and Si$_{0.14}$Ge$_{0.80}$Sn$_{0.06}$ on SOI in b), d), and f). The EDXS-based maps in a) and b) are the superimposed element distributions of Si (red), O (green), Ge (blue), and Sn (magenta). The HR-TEM images in c) and d) show from bottom to top: burried SiO$_2$, Si, and the grown Ge$_{0.94}$Sn$_{0.06}$ or Si$_{0.14}$Ge$_{0.80}$Sn$_{0.06}$ layer. The white dashed rectangles in c) and d) correspond to the enlarged images in e) and f).

The microstructure after PLA with 0.20 J cm$^{-2}$ (Ge$_{0.94}$Sn$_{0.06}$) or 0.25 J cm$^{-2}$ (Si$_{0.14}$Ge$_{0.80}$Sn$_{0.06}$) is shown in Fig. 2. Both materials have a highly improved crystal structure compared to their as-grown states. The $Ge_{1-x}Sn_x$ sample is fully crystalline after PLA at 0.20 J cm$^{-2}$ but shows local inhomogeneities in the $Ge_{1-x}Sn_x$ layer thickness (Fig. 2 c) and e)). Additionally, element redistributions are observed, which indicate a locally overheated $Ge_{1-x}Sn_x$ layer (Fig. 2 a)). The $Si_{0.14}Ge_{0.80}Sn_{0.06}$ sample PLA-treated at 0.25 J cm$^{-2}$ also shows a single-crystalline structure with only few defects (see (Fig. 2 d) and f)). The EDXS-based element distribution maps in Fig. 2 a) and b) suggest an out-diffusion of Sn and the formation of a thick Sn-rich oxide at the sample surface. On the other hand, any formation of Sn clusters or filaments was not observed after PLA with low energy densities, which is significantly different compared to pulse laser melting results of $Ge_{1-x}Sn_x$ [22-24].

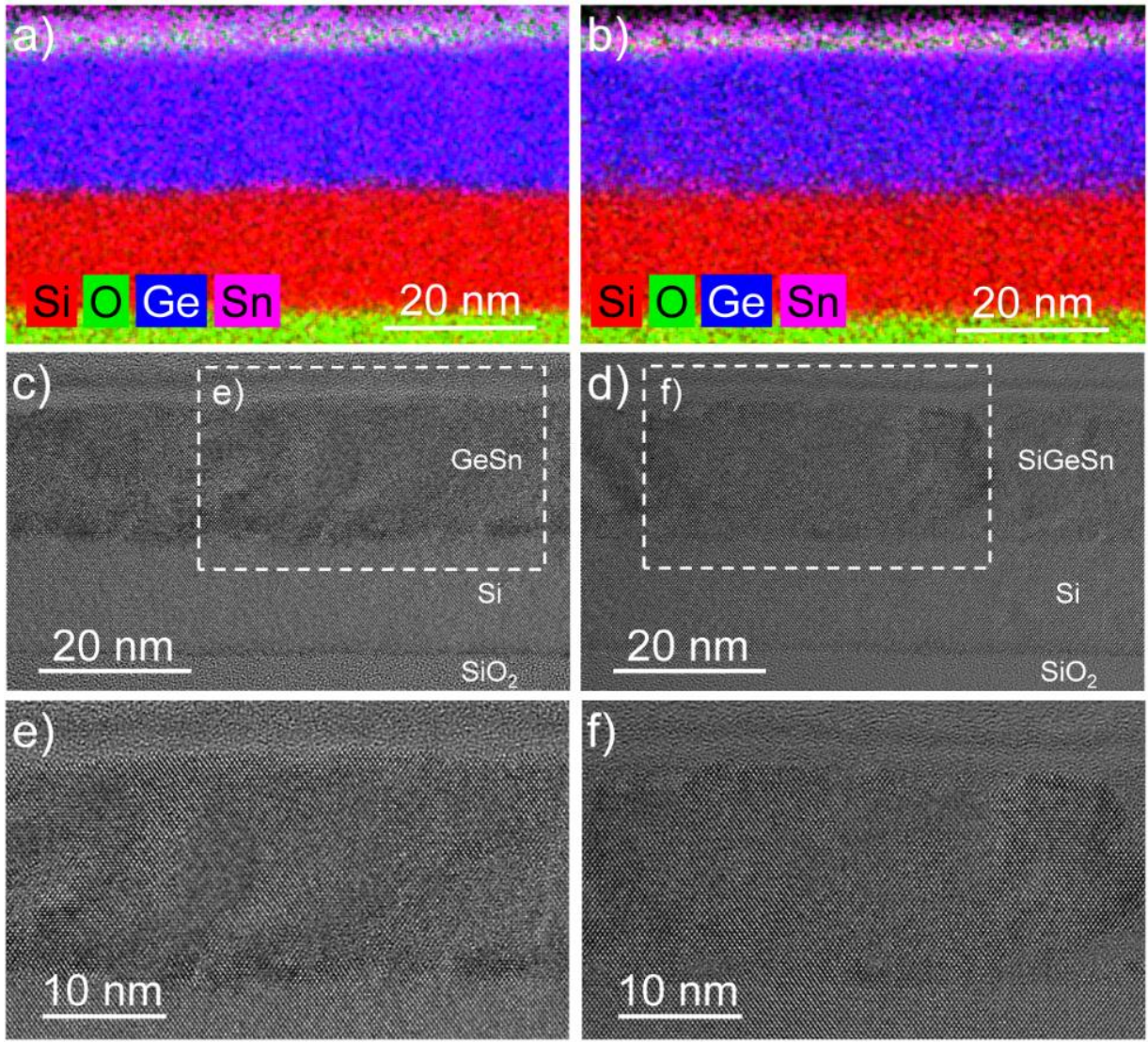

Fig. 2: TEM-based analysis of $Ge_{0.94}Sn_{0.06}$ on SOI after PLA at 0.20 J cm$^{-2}$ in a), c), and e) and $Si_{0.14}Ge_{0.80}Sn_{0.06}$ on SOI after PLA at 0.25 J cm$^{-2}$ in b), d), and f). The EDXS-based maps in a) and b) are the superimposed element distributions of Si (red), O (green), Ge (blue), and Sn (magenta). The white dashed rectangles in c) and d) correspond to the enlarged images in e) and f).

Since the out-diffusion of Sn is an undesired process, the PLA energy densities $E_d$ were systematically reduced, and the crystal structure evolution was investigated by XRD and RBS. Fig. 3 and Fig. 4 show 2 2 4 RSMs for $Ge_{0.94}Sn_{0.06}$ and $Si_{0.14}Ge_{0.80}Sn_{0.06}$ samples annealed at different conditions, respectively. The as-grown state of Ge$_{0.94}$Sn$_{0.06}$ in Fig. 3 a) has a barely noticeable GeSn 2 2 4 reflection located close to the strain relaxation line at $q_z/2\pi \approx 7.0$ nm$^{-1}$. This weak diffraction is related to the mostly amorphous nature of the Ge$_{0.94}$Sn$_{0.06}$ layer. After PLA, the peak intensity increases significantly. Furthermore, the location of the reflection on the black dashed strain relaxation line confirms that the Ge$_{1-x}$Sn$_x$ layer is almost strain-relaxed on the SOI substrate. Converting the reciprocal in-plane $q_x$ and out-of plane $q_z$ peak lattice parameters into the real space in-plane $a_{//}$ and out-of-plane $a_\perp$ lattice parameters by Eq. 1 and Eq. 2 leads to a relaxed lattice parameter $a_\perp \approx a_{//} \approx 0.57$ nm.

$$a_\perp = 4\,\frac{2\pi}{q_z} \qquad \text{Eq. 1}$$

$$a_{II} = \sqrt{8}\,\frac{2\pi}{q_x} \qquad \text{Eq. 2}$$

Applying the bowing-parameter-corrected Vegard's law (see Eq. 3) [25], where $a_{Ge}$ is the lattice parameter of pure Ge, $\Delta_{GeSn}$ is the difference of the elemental Ge and Sn lattice parameters, and $\Psi_{GeSn} = 0.016$ nm [25] is the bowing parameter, allows to approximate the Sn concentration in Ge$_{1-x}$Sn$_x$ to x ≈ 5.97 at.% after PLA. Since the Sb incorporation was not taken into account, the given Sn concentration should be taken as an estimate.

$$a_{Ge_{1-x}Sn_x} = a_{Ge} + \Delta_{GeSn}\,x + \Psi_{GeSn}\,x\,(1-x) \qquad \text{Eq. 3}$$

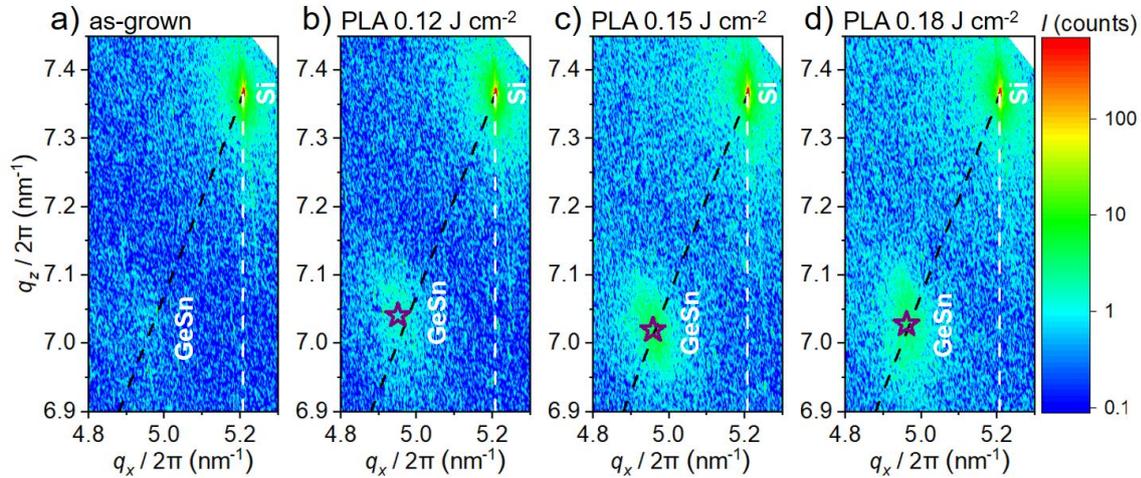

Fig. 3: 2 2 4 RSM of Ge$_{1-x}$Sn$_x$ on SOI in the as-grown state a) and after PLA with $E_d$ = 0.12 J cm$^{-2}$ b), 0.15 J cm$^{-2}$ c), and 0.18 J cm$^{-2}$ d). The black dashed line is the strain relaxation line, and the vertical white dashed line corresponds to the fully pseudomorphically grown state. The fitted maxima of the GeSn 2 2 4 reflections are indicated by the purple stars.

The 2 2 4 reflection of the as-grown Si$_{0.14}$Ge$_{0.80}$Sn$_{0.06}$ in Fig. 4 a) is located between the pseudomorphic and the strain relaxation lines. This suggests a strain distribution across the layer thickness from partially compressive-strained at the Si$_{0.14}$Ge$_{0.80}$Sn$_{0.06}$/SOI interface toward almost relaxed close to the surface. After PLA with 0.15 J cm$^{-2}$, the SiGeSn 2 2 4 reflection appears to be elongated in the $q_z$ direction. While the portion with $q_z/2\pi \approx 7.0$ nm$^{-1}$ reminds in location and intensity of the as-grown state, the contribution with the higher intensity at $q_z/2\pi \approx 7.1$ nm$^{-1}$ appears to be slightly tensile-strained. This is related to a separation of the $Si_{1-x-y}Ge_ySn_x$ layer into an as-grown-like bottom layer and a slightly tensile-strained top-layer. This tensile strain might be caused by the thermal expansion coefficient differences between the alloy and the substrate, and the large temperature gradient during PLA. Tensile strain was observed earlier for Ge on Si [21, 26] or $Ge_{1-x}Sn_x$ on Si [27] after thermal treatments. In the case of $Si_{0.14}Ge_{0.80}Sn_{0.06}$ annealed with $E_d$ = 0.20 J cm$^{-2}$ in Fig. 4 c), both reflections merged, and the SiGeSn 2 2 4 reflection has a similar $q_x/q_z$ position as in the as-grown state. This is related to the increase in effectively heated volume with the increasing $E_d$ of the PLA laser. PLA with $E_d$ = 0.25 J cm$^{-2}$ causes a shift of the SiGeSn 2 2 4 reflection parallel to the strain relaxation line towards the Si 2 2 4 substrate reflection (Fig. 4 d)). This is a clear evidence of the out-diffusion of a significant amount of larger atoms since the strain conditions remain almost the same, and it correlates with the lower Sn contrast in the EDX-based element distribution map in Fig. 2 b).

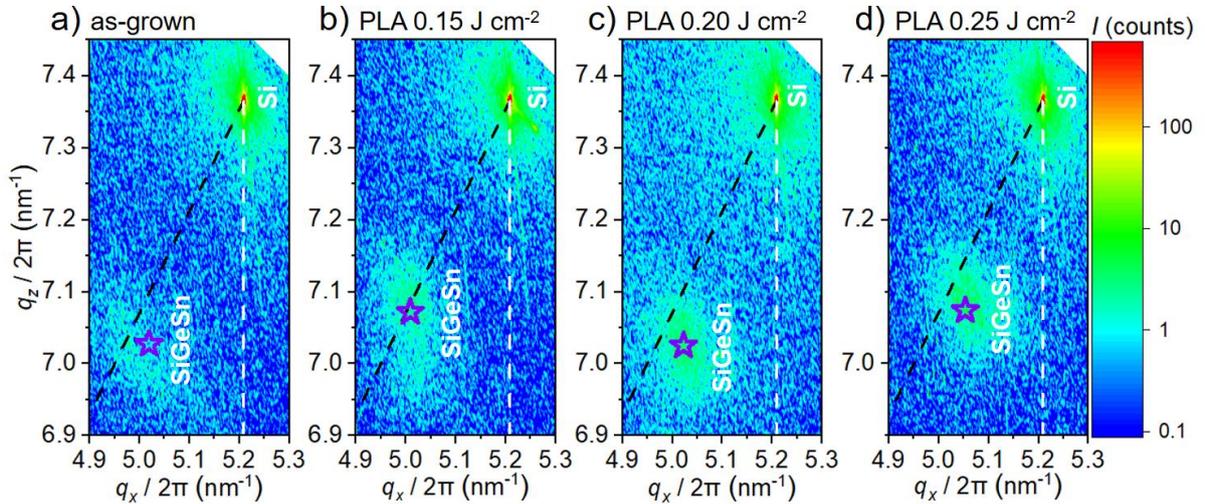

Fig. 4: 2 2 4 XRD-RSM of $Si_{0.14}Ge_{0.80}Sn_{0.06}$ on SOI in the as-grown state a) and after PLA with $E_d$ = 0.15 J cm$^{-2}$ b), 0.20 J cm$^{-2}$ c), and 0.25 J cm$^{-2}$ d). The black dashed line is the strain relaxation line, and the vertical white dashed line corresponds to the fully pseudomorphically grown state. The fitted maximum of the SiGeSn 2 2 4 reflection is indicated by the center of the purple star.

The RBS-R/C results of the $Ge_{1-x}Sn_x$ and $Si_{1-x-y}Ge_ySn_x$ on SOI before and after annealing are shown in Fig. 5 and Fig. 6, respectively. The obtained spectra contain Sn superimposed with Sb (1460 - 1510 keV), Ge (1325 – 1390 keV), and Si (< 1000 keV). Since the Sb concentration is only about 0.1 at. %, the signal in the spectra is mainly related to Sn. The Si contribution of $Si_{0.14}Ge_{0.80}Sn_{0.06}$, located between 950 – 1000 keV, is merged with the pure Si signal of the SOI layer and the high-energy tail of the burried $SiO_2$.

The RBS-R spectra of the as-grown $Ge_{0.94}Sn_{0.06}$ and $Si_{0.14}Ge_{0.80}Sn_{0.06}$ states were fitted by SIMNRA and confirmed the expected layer compositions and thicknesses.

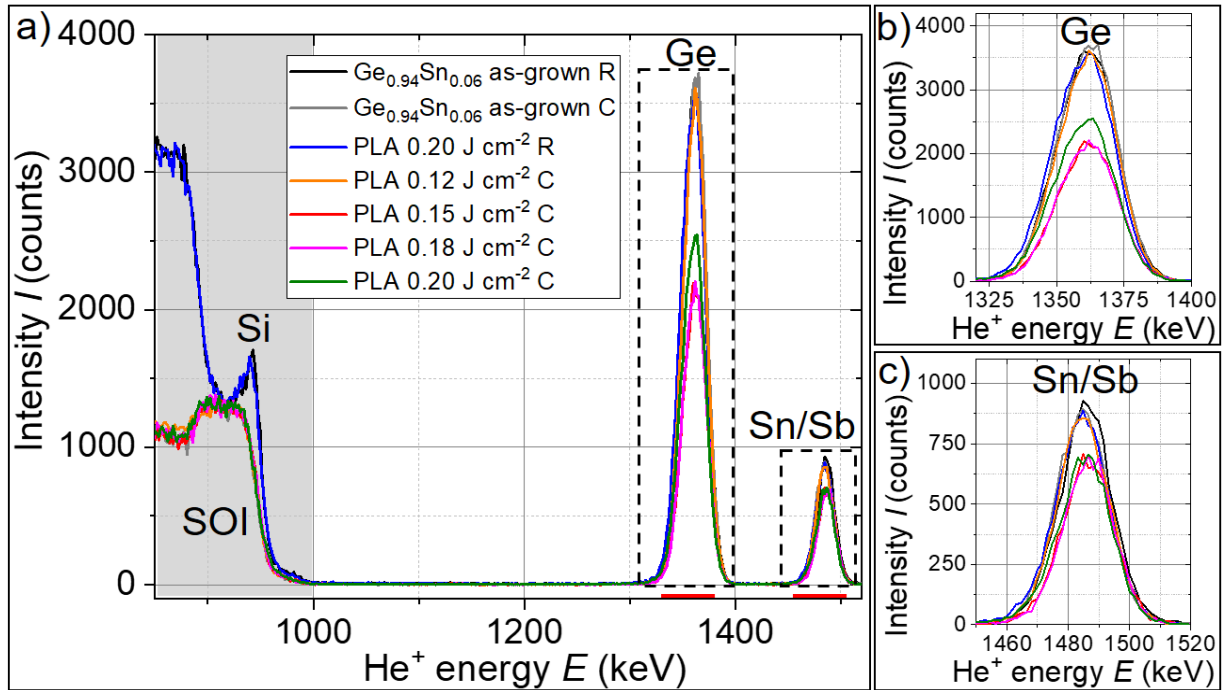

Fig. 5: RBS-R/C results of Ge$_{0.94}$Sn$_{0.06}$ on SOI in the as-grown state and after PLA with 0.12, 0.15, 0.18, and 0.20 J cm$^{-2}$ a). The marked dashed windows in a) show the enlargements of the Ge b) and Sn/Sb c) contributions of Ge$_{1-x}$Sn$_x$. The spectra of Sn and Sb are superimposed, and the contributions from the SOI substrate are indicated by the grey background.

The RBS-C spectrum of the Ge$_{0.94}$Sn$_{0.06}$ as-grown state in Fig. 5 is similar to its RBS-R counterpart since the layer contains amorphous inclusions. Analog channeling results were also obtained for the mildly treated PLA 0.12 J cm$^{-2}$ sample. Channeling effects in the Si, Ge, and Sn/Sb signals can be observed after PLA with 0.15, 0.18, and 0.20 J cm$^{-2}$.

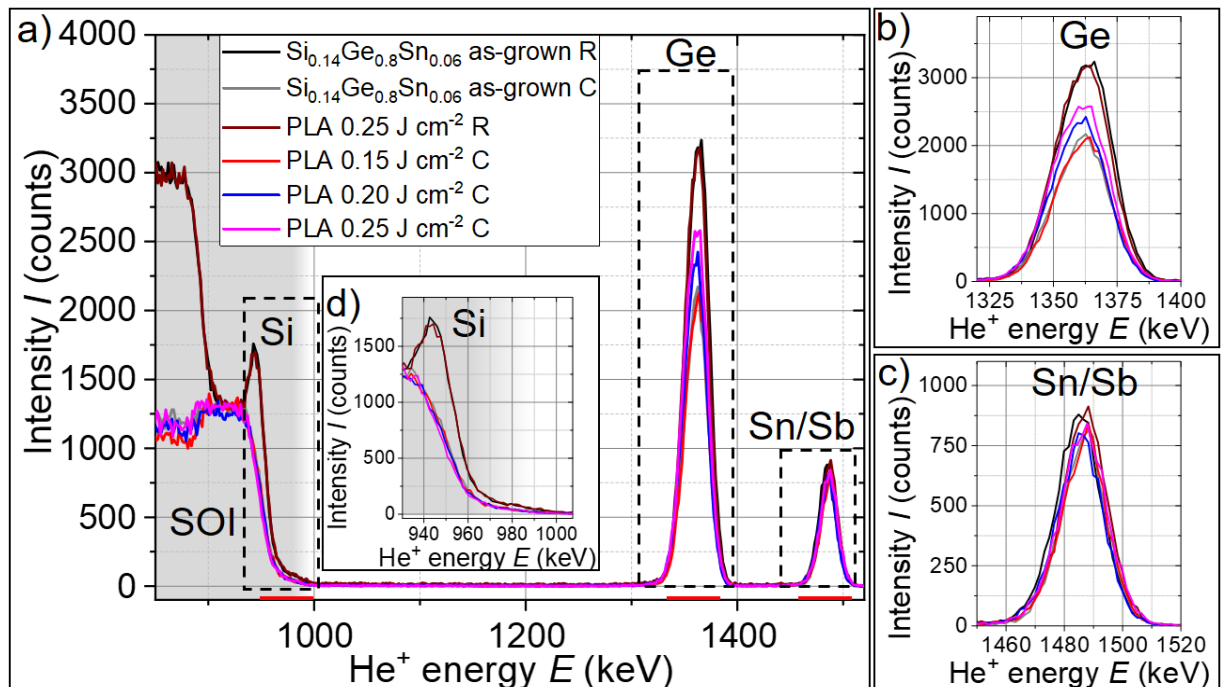

Fig. 6: RBS-R/C results of Si$_{0.14}$Ge$_{0.80}$Sn$_{0.06}$ on SOI in the as-grown state and after PLA with 0.15 J cm$^{-2}$, 0.20 J cm$^{-2}$, and 0.25 J cm$^{-2}$ a). The marked dashed windows in a) show the enlargements of the Ge b), the superimposed Sn/Sb c), and the Si d) contributions. The contributions from the SOI substrate, indicated by the grey background, superimpose the Si contribution from the Si$_{1-x-y}$Ge$_y$Sn$_x$ layer.

The occurrence of channeling in Si, Ge, and Sn/Sb in the $Si_{0.14}Ge_{0.80}Sn_{0.06}$ as-grown state in Fig. 6 confirms the epitaxial growth of $Si_{1-x-y}Ge_ySn_x$ on the SOI wafer. This matches the TEM results in Fig. 1 d) and f). PLA of $Si_{0.14}Ge_{0.80}Sn_{0.06}$ with $E_d \geq 0.15$ J cm$^{-2}$ causes a slight de-channeling compared to the $Si_{0.14}Ge_{0.80}Sn_{0.06}$ as-grown state.

The RBS-R/C spectra were quantitatively analyzed by calculating the channeling yield $\chi$ for Si, Ge, and Sn with Eq. 4, and the substitutional fraction $\xi$ of Sn/Sb on Ge lattice sites $\xi_{Sn/Sb,Ge}$ by Eq. 5.

$$\chi = \frac{A_C}{A_R} \qquad \text{Eq. 4}$$

$A_c$ is the integrated area under the channeling curve, and $A_R$ is the integrated area under the random curve.

$$\xi_{Sn/Sb,Ge} = \frac{(1-\chi_{Sn/Sb})}{(1-\chi_{Ge})} \qquad \text{Eq. 5}$$

In general, the obtained $\chi_{Si}$, $\chi_{Ge}$, and $\chi_{Sn/Sb}$ of the fabricated $Ge_{1-x}Sn_x$ and $Si_{1-x-y}Ge_ySn_x$ in Table 1 are relatively high because of i) the observed defects in Fig. 1 and Fig. 2 can displace the atomic lattice positions, ii) the superposition of crystal channeling with de-channeling events from surface defects, and iii) the formation of the thicker $Ge_{1-x}Sn_x$- or $Si_{1-x-y}Ge_ySn_x$-oxide layer after PLA (see Fig. 2 a) and b)). The presence of this oxide layer might be avoidable if the PLA process is performed under an inert gas atmosphere or *in situ* in an MBE cluster tool. The $\xi_{Sn/Sb,Si}$ calculation results for the incorporation of Sn/Sb on Si substitutional sites are not presented since the superposition of the Si signal with the top Si and $SiO_2$ tail regions of the SOI causes errors in the $\chi_{Si}$ result.

Table 1: RBS analysis results of $Ge_{0.94}Sn_{0.06}$ and $Si_{0.14}Ge_{0.80}Sn_{0.06}$ layers on SOI in the as-grown and PLA-treated states. The minimum channeling yield for Si $\chi_{Si}$, Ge $\chi_{Ge}$, and Sn/Sb $\chi_{Sn/Sb}$, as well as the substitutional fraction of Sn/Sb on Ge lattice sites $\xi_{Sn/Sb,Ge}$ is calculated by Eq. 4 and Eq. 5. The integration intervals are marked with red horizontal section lines in Fig. 5 a) and Fig. 6 a) and are between 950 and 1000 keV for Si, 1340 and 1390 keV for Ge, and 1465 and 1515 keV for Sn/Sb.

| Sample | $\chi_{Si}$ (%) | $\chi_{Ge}$ (%) | $\chi_{Sn/Sb}$ (%) | $\xi_{Sn/Sb,Ge}$ (%) |
|---|---|---|---|---|
| $Ge_{0.94}Sn_{0.06}$ as-grown | - | 100.0 | 97.8 | 0 |
| $Ge_{1-x}Sn_x$ PLA 0.15 J cm$^{-2}$ | - | 60.6 | 76.5 | 59.6 |
| $Ge_{1-x}Sn_x$ PLA 0.18 J cm$^{-2}$ | - | 58.9 | 72.7 | 66.4 |
| $Ge_{1-x}Sn_x$ PLA 0.20 J cm$^{-2}$ | - | 71.2 | 79.6 | 70.7 |
| $Si_{0.14}Ge_{0.80}Sn_{0.06}$ as-grown | 48.2 | 58.8 | 80.5 | 47.2 |
| $Si_{1-x-y}Ge_ySn_x$ PLA 0.15 J cm$^{-2}$ | 52.2 | 63.1 | 84.3 | 42.7 |
| $Si_{1-x-y}Ge_ySn_x$ PLA 0.20 J cm$^{-2}$ | 53.4 | 71.6 | 87.4 | 44.2 |
| $Si_{1-x-y}Ge_ySn_x$ PLA 0.25 J cm$^{-2}$ | 52.4 | 79.0 | 93.7 | 30.0 |

The appearance of channeling in the $Ge_{1-x}Sn_x$ microstructure and the SOI beneath suggest epitaxial recrystallization or regrowth during the PLA process. The lowest channeling yields in Table 1 were achieved after PLA with 0.15 and 0.18 J cm$^{-2}$ with $\chi_{Ge} \approx 60\%$ and $\chi_{Sn/Sb} \approx 75\%$. Additionally, the occupation of Sn/Sb on Ge lattice sites with $\xi_{Sn/Sb,Ge}$ = 59.6 % and 66.4 % also suggests a reasonable amount of incorporated/activated Sn/Sb atoms on substitutional lattice sites. However, further increasing the PLA to $E_d$ = 0.20 J cm$^{-2}$ causes de-channeling in Ge and Sn/Sb. For Si$_{0.14}$Ge$_{0.80}$Sn$_{0.06}$, the best channeling properties were observed for the as-grown state. However, the direct comparison of Fig. 1 f) and Fig. 2 f) clearly shows an improved crystal structure. Hence, the channeling in Ge$_{1-x}$Sn$_x$ and Si$_{1-x-y}$Ge$_y$Sn$_x$ is most likely systematically overestimated due to the earlier-mentioned de-channeling effects (oxide layer, and surface defects).

The dopant concentration and distribution in dependence of the PLA $E_d$ were investigated by SIMS and Hall-effect measurements. The SIMS results in Fig. 7 and Fig. 8 show the element depth distribution before and after annealing. Both *in situ*-doped as-grown Ge$_{0.94}$Sn$_{0.06}$ and Si$_{0.14}$Ge$_{0.80}$Sn$_{0.06}$ on SOI samples have a relatively homogeneous Sb intensity profile despite the surface-related amplitude (see Fig. 7). Unfortunately, the Sn concentration is slightly reduced after PLA.

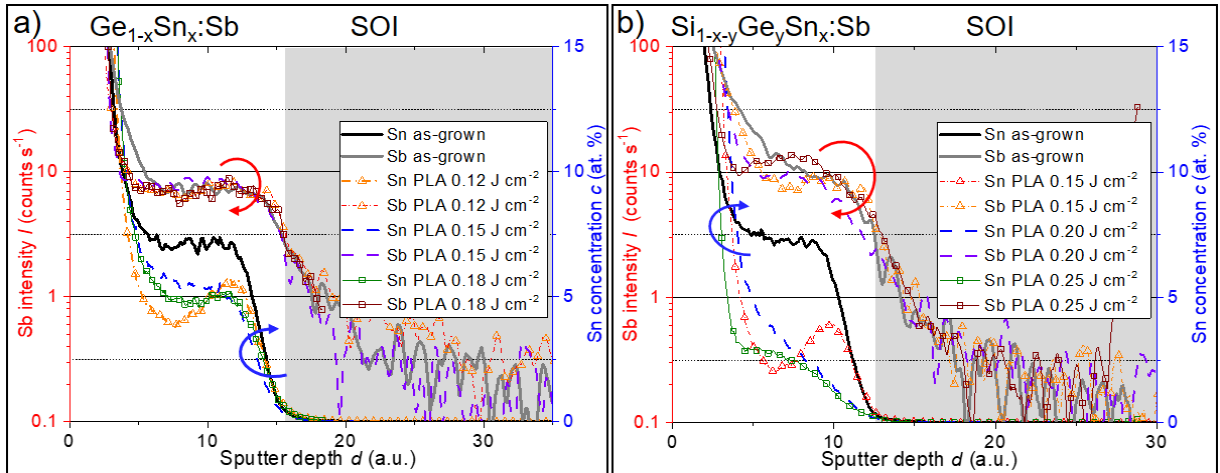

Fig. 7. Sb and Sn depth distributions of the Ge$_{1-x}$Sn$_x$ on SOI a) and Si$_{1-x-y}$Ge$_y$Sn$_x$ on SOI b) in the as-grown state and after PLA measured by TOF-SIMS. The Si layer beneath the alloy is colored in grey. The sputter depth was aligned to the Ge and Si crossover at the alloy/SOI interface shown in Fig. 8. A relative error within the given Sn concentration of ±1 at. % is expected.

The Si and Ge profiles of the Ge$_{1-x}$Sn$_x$ samples in Fig. 8 a) are relatively homogenous, but the Si$_{1-x-y}$Ge$_y$Sn$_x$ layers shows a redistribution of Ge and Si after PLA in Fig. 8 b). Furthermore, the PLA 0.15 J cm$^{-2}$ state shows a small kink close to the SOI interface. This kink is related to the almost PLA-unaffected Si$_{1-x-y}$Ge$_y$Sn$_x$ layer, which correlates with the weak intensity signal below the main SiGeSn 2 2 4) reflection in Fig. 4 b). The sample annealed with 0.20 J cm$^{-2}$ shows a relatively constant Ge concentration while Sn increases and Si decreases toward the surface. The slightly varying chemical composition across the layer thickness and the different crystal qualities influence the sputter yield and might reduce the accuracy in the quantitative SIMS analysis. This could be the reason for the slightly overestimated Sn concentration in the as-grown states in Fig. 7. After annealing with 0.15 and 0.25 J cm$^{-2}$, a redistribution of Si toward the SOI interface and a Ge diffusion towards the surface is visible.

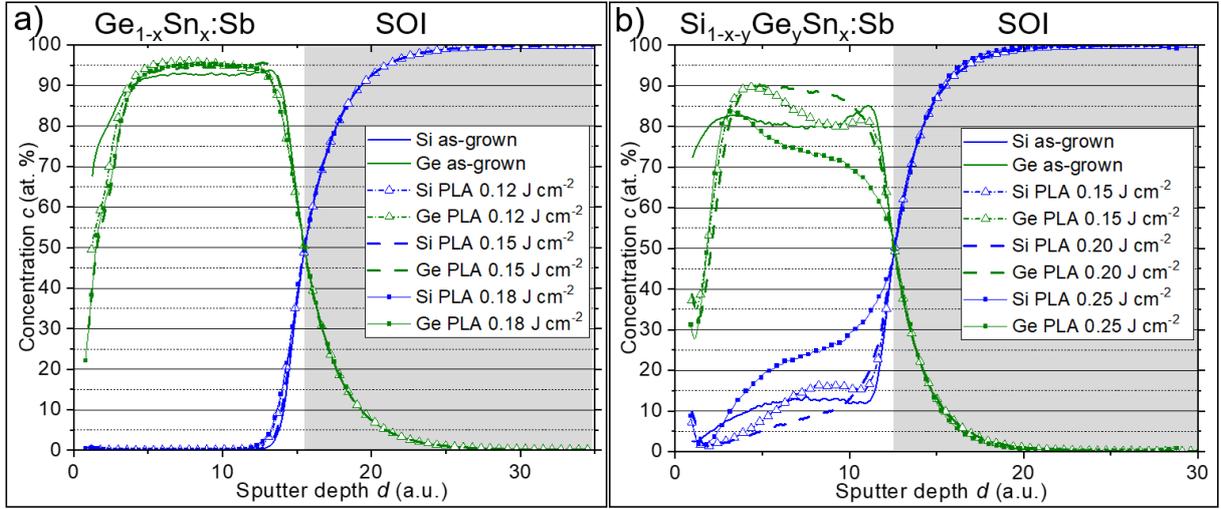

Fig. 8. Ge and Si depth distribution of the Ge$_{1-x}$Sn$_x$ on SOI a) and Si$_{1-x-y}$Ge$_y$Sn$_x$ on SOI b) in the as-grown state and after PLA measured by TOF-SIMS. The Si layer beneath the alloy is colored in grey. The sputter depth was aligned to the Ge and Si crossover at the alloy/SOI interface.

The Hall-effect results of the Ge$_{1-x}$Sn$_x$ and Si$_{1-x-y}$Ge$_y$Sn$_x$ layers before and after post-growth PLA are shown in Fig. 9. The carrier concentrations in the as-grown states are 1.8 x 10$^{19}$ cm$^{-3}$ (Ge$_{0.94}$Sn$_{0.06}$) and 5.1 x 10$^{17}$ cm$^{-3}$ (Si$_{0.14}$Ge$_{0.80}$Sn$_{0.06}$). After PLA with $E_d \geq$ 0.15 J cm$^{-2}$, a significant amount of Sb could be activated in both alloys. In the case of Ge$_{1-x}$Sn$_x$, active carrier concentrations of 3.9 x 10$^{19}$ cm$^{-3}$ (0.15 J cm$^{-2}$) and 4.2 x 10$^{19}$ (0.18 J cm$^{-2}$) were determined, which are close to the targeted absolute Sb concentration of 5 x 10$^{19}$ cm$^{-3}$. In the Si$_{1-x-y}$Ge$_y$Sn$_x$ case, $n_{e^-}$ could be increased up to 2.3 x 10$^{19}$ cm$^{-3}$ ($E_d$ = 0.15 J cm$^{-2}$). However, a too high PLA energy density reduces the active carrier concentration, as shown for Si$_{1-x-y}$Ge$_y$Sn$_x$ with $E_d$ = 0.25 J cm$^{-2}$.

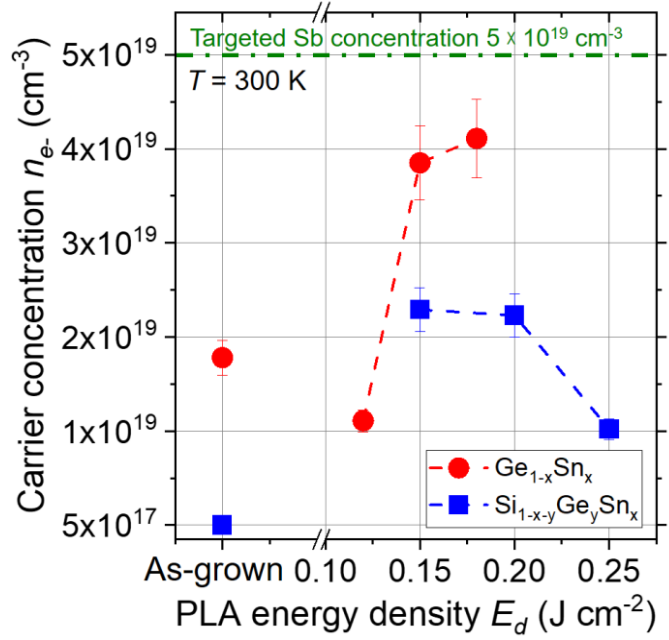

Fig. 9: Active electron concentration $n_{e^-}$ of $Ge_{1-x}Sn_x$ and $Si_{1-x-y}Ge_ySn_x$ determined by Hall-effect measurements at 300 K of the as-grown states ($E_d$ = 0 J cm$^{-2}$) and after post-growth PLA with different energy densities $E_d$.

## IV. Conclusion

The studied $Ge_{0.94}Sn_{0.06}$ and $Si_{0.14}Ge_{0.80}Sn_{0.06}$ films grown on SOI substrates are almost strain-relaxed and contain many defects since the layer thickness exceeds the critical thickness for plastic strain relaxation. Partially replacing Ge by Si reduces the lattice mismatch between the alloy and the Si substrate from about 5.3 % ($Ge_{0.94}Sn_{0.06}$) to about 1.9 % ($Si_{0.14}Ge_{0.80}Sn_{0.06}$) and helps to improve the crystal structure. Post-growth PLA improves the layer quality significantly and activates up to 80% of the Sb-donors. However, annealing under ambient conditions generates a Sn-rich oxide on the surface and slightly redistributes Si, Ge, and Sn.

Overall, it can be concluded that the $Ge_{0.94}Sn_{0.06}$ and $Si_{0.14}Ge_{0.80}Sn_{0.06}$ grown directly on SOI layers contain many defects, which is a challenge for their application as an active component in opto- or nanoelectronic devices. Hence, it is suggested to use a post-growth treatment to mediate between the large lattice mismatch of the alloy and the substrate. Such a treatment should be performed in an inert atmosphere or vacuum to avoid oxide formation. The results show that PLA of Sn-containing group-IV alloys on SOI wafers is an efficient way to fabricate $Ge_{1-x}Sn_x$ and $Si_{1-x-y}Ge_ySn_x$ alloys compatible with CMOS technology on an insulating platform.

## V. Supplementary materials

The supplements support the presented XRD findings. Supplementary materials A discusses a possible tilt between the top silicon and the Si carrier substrate of the SOI wafer. Supplementary materials B shows and explains additional 0 0 4 HR-XRD results, which support the discussed 2 2 4 XRD-RSM data.

## ACKNOWLEDGMENTS

This work was partially supported by the Bundesministerium für Bildung und Forschung (BMBF) under the project "ForMikro": Group IV heterostructures for high-performance nanoelectronic devices (SiGeSn NanoFETs) (Project-ID: 16ES1075). We gratefully acknowledge the HZDR Ion Beam Centre for their support with RBS. The authors thank Annette Kunz for TEM specimen preparation. Furthermore, the use of the HZDR Ion Beam Center TEM facilities and the funding of TEM Talos by the Bundesministerium für Bildung und Forschung (BMBF), Grant No. 03SF0451, in the framework of HEMCP are acknowledged. We thank Dr. Olav Hellwig and his group at HZDR for providing access to the X-ray diffractometer.

## DATA AVAILABILITY

The data that support the findings of this study are available from the corresponding author upon reasonable request.

# Supplementary materials A: SOI wafer tilt

The knowledge of the tilt angle between the thick Si carrier substrate and the $Si_{1-x-y}Ge_ySn_x$ layer is necessary to analyze the 2 2 4 RSM results correctly. In the case of epitaxial growth on a single-crystalline substrate, an almost preserved lattice orientation compared to the substrate can be assumed. However, the 20 nm-thick SOI layer, which was used as the seed for the epitaxial growth of $Ge_{0.94}Sn_{0.06}$ or $Si_{0.14}Ge_{0.80}Sn_{0.06}$, can be tilted relative to the about 750 µm-thick carrier substrate. The tilt was investigated with RSMs around the 0 0 4 reflection of the film after a rocking curve alignment on the Si carrier substrate reflection. The symmetrical 0 0 4 reflection was used since $q_x/2\pi$ must be located at 0 nm$^{-1}$ if the layer is epitaxially grown. Any deviation of the $Si_{1-x-y}Ge_ySn_x$ reflection from $q_x/2\pi = 0$ nm$^{-1}$ can be understood as a layer tilt. The reciprocal lattice parameters $q_x$ and $q_z$ of the substrate´s 0 0 4 and the $Si_{1-x-y}Ge_ySn_x$ 0 0 4 layer reflections were determined by fitting their positions, as shown in Fig. A 1. $q_x$ is the in-plane component and $q_z$ the out-of-plane component of the reciprocal scattering vector Q.

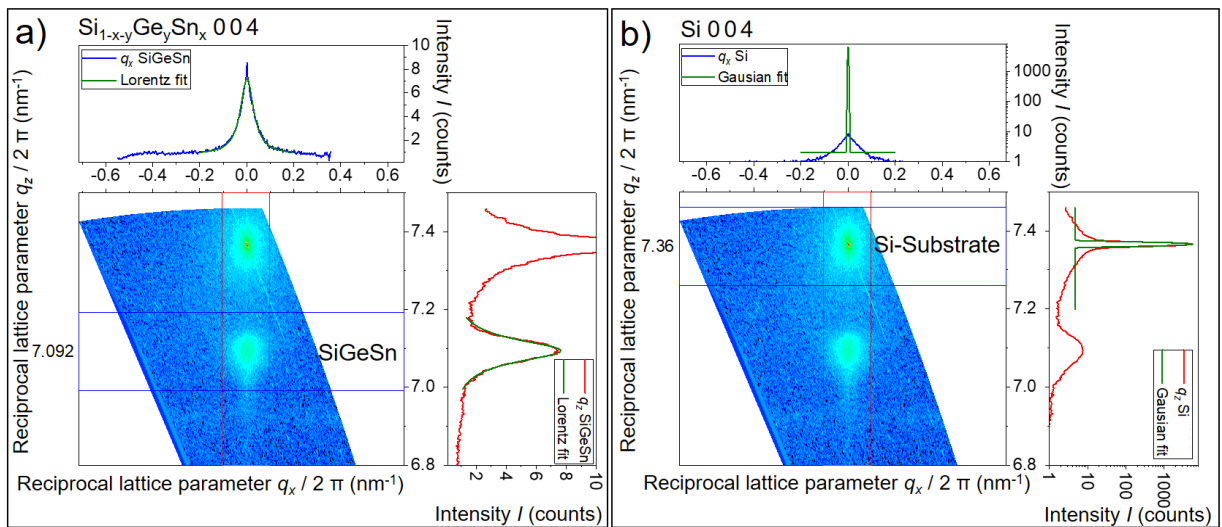

Fig. A 1: Fitting procedure of $q_x$ and $q_z$ based on an RSM around the 0 0 4 reflection of the epitaxially grown $Si_{1-x-y}Ge_ySn_x$ layer a) and the 0 0 4 peak of the Si carrier substrate b). The SiGeSn peak position was fitted with a Lorentzian function, and the Si substrate was fitted with a Gaussian function. For both cases, the adjusted $R^2$ was above 98 %. The Lorentzian function was used due to a slight asymmetryy in the SiGeSn peak shape.

Afterwards, the tilt angle $δ$ was calculated using Eq. A 1. The calculated tilt between the Si substrate and the $Ge_{1-x}Sn_x$ and $Si_{1-x-y}Ge_ySn_x$ alloys was below 1 x 10$^{-2}$° for all investigated samples. This tilt is small enough to be neglected. Therefore, it can be concluded that all shifts of the asymmetrical 2 2 4 reflection in Fig. 3 and Fig. 4 are caused by changes in the $Ge_{1-x}Sn_x$ or $Si_{1-x-y}Ge_ySn_x$ alloy lattice parameters, respectively.

$$\delta = \arctan\left(\frac{q_{x,Si\ 0\ 0\ 4}}{q_{z,Si\ 0\ 0\ 4}}\right) - \arctan\left(\frac{q_{x,SiGeSn\ 0\ 0\ 4}}{q_{z,SiGeSn\ 0\ 0\ 4}}\right) \qquad \text{Eq. A 1}$$

# Supplementary materials B: 0 0 4 HR-XRD

HR-XRD $\theta$-$2\theta$ scans of the 0 0 4 reflection were performed to confirm the 2 2 4 RSM findings. The obtained diffractograms contain the $Ge_{1-x}Sn_x$ 0 0 4 reflection in Fig. B 1 a) and $Si_{1-x-y}Ge_ySn_x$ 0 0 4 reflection in Fig. B 1 b) reflections between 63° and 68° and the not-shown Si 0 0 4 substrate reflection at 69.14°. The substrate lattice was used to align the measurement setup, and a general substrate tilt was excluded, as shown in Appendix A.

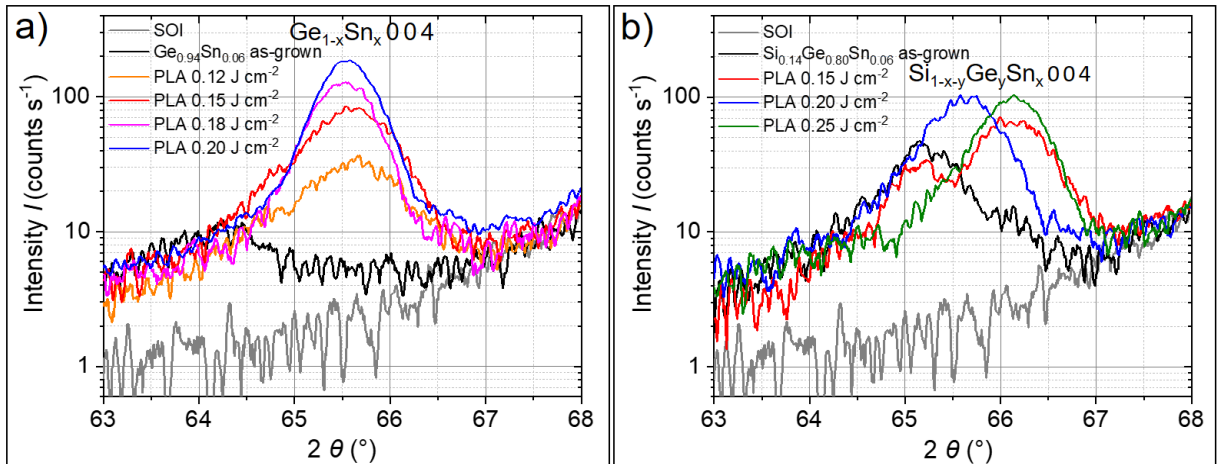

Fig. B 1: HR-XRD $\theta$-$2\theta$ scans around the 0 0 4 reflection of the $Ge_{1-x}Sn_x$ on SOI a) and $Si_{1-x-y}Ge_ySn_x$ on SOI b) in the as-grown state and after PLA with varying energy densities.

Both as-grown states of the investigated alloys have relatively low reflection intensities in Fig. B 1. After PLA, a shift of the $Ge_{1-x}Sn_x$ 0 0 4 and $Si_{1-x-y}Ge_ySn_x$ 0 0 4 reflection toward larger angles and an increased reflection intensity can be observed. The higher intensities correlate partially with the 2 2 4 RSM results in Fig. 3 and Fig. 4 and can be understood as an improvement of the lattice compared to the as-grown state. This is also reflected in the reduced full width at half maximum of the $Ge_{1-x}Sn_x$ 0 0 4 reflection. Furthermore, a significant shift towards higher diffraction angles is observed after PLA for both materials in Fig. B 1. This indicates a reduced out-of-plane lattice parameter, which is most likely caused by the out-diffusion of larger atoms. Next to the main high-intensity 0 0 4 reflections, an additional small peak at around 65° can be observed after PLA with $E_d$ = 0.15 J cm$^{-2}$ for both materials. This can be related to a layer separation into a slightly annealed as-grown-like bottom layer and the highly PLA-affected top layer. In the case of $Si_{1-x-y}Ge_ySn_x$ after PLA at 0.20 J cm$^{-2}$, the two reflections merge, and the maximum shifts only slightly toward lower diffraction angles compared to the main reflection of $Si_{1-x-y}Ge_ySn_x$ after PLA with 0.15 and 0.25 J cm$^{-2}$.

# References

[1] Y. Liu, J. Yan, M. Liu, H. Wang, Q. Zhang, B. Zhao, C. Zhang, B. Cheng, Y. Hao, G. Han, Mobility enhancement in undoped Ge0.92Sn0.08 quantum well p-channel metal-oxide-semiconductor field-effect transistor fabricated on (111)-oriented substrate, Semiconductor Science and Technology 29 (2014) 115027. https://doi.org/10.1088/0268-1242/29/11/115027

[2] S. Gupta, B. Magyari-Köpe, Y. Nishi, K.C. Saraswat, Achieving direct band gap in germanium through integration of Sn alloying and external strain, Journal of Applied Physics 113 (2013) 073707. https://doi.org/10.1063/1.4792649

[3] J.D. Sau, M.L. Cohen, Possibility of increased mobility in Ge-Sn alloy system, Phys Rev B 75 (2007) 045208. https://doi.org/10.1103/PhysRevB.75.045208

[4] B. Mukhopadhyay, G. Sen, R. Basu, S. Mukhopadhyay, P.K. Basu, Prediction of Large Enhancement of Electron Mobility in Direct Gap Ge1−xSnx Alloy, physica status solidi B 254 (2017) 1700244. https://doi.org/10.1002/pssb.201700244

[5] J. Kaur, R. Basu, A.K. Sharma, Design and Analysis of Si1-x-yGeySnx-Si1-xGex Alloy Based Solar Cell Emphasizing on Ge Composition 15%, Silicon 15 (2022) 397–404. https://doi.org/10.1007/s12633-022-02025-7

[6] D. Lei, K.H. Lee, Y.-C. Huang, W. Wang, S. Masudy-Panah, S. Yadav, A. Kumar, Y. Dong, Y. Kang, S. Xu, Y. Wu, C.S. Tan, X. Gong, Y.-C. Yeo, Germanium-Tin (GeSn) P-Channel Fin Field-Effect Transistor Fabricated on a Novel GeSn-on-Insulator Substrate, IEEE Transactions on Electron Devices 65 (2018) 3754-3761. https://doi.org/10.1109/TED.2018.2856738

[7] S. Wirths, D. Buca, S. Mantl, Si–Ge–Sn alloys: From growth to applications, Progress in Crystal Growth and Characterization of Materials 62 (2016) 1-39. https://doi.org/10.1016/j.pcrysgrow.2015.11.001

[8] D. Schwarz, H.S. Funk, M. Oehme, J. Schulze, Alloy Stability of Ge1−xSnx with Sn Concentrations up to 17% Utilizing Low-Temperature Molecular Beam Epitaxy, Journal of Electronic Materials 49 (2020) 5154-5160. https://doi.org/10.1007/s11664-020-08188-6

[9] S. Zaima, O. Nakatsuka, N. Taoka, M. Kurosawa, W. Takeuchi, M. Sakashita, Growth and applications of GeSn-related group-IV semiconductor materials, Sci Technol Adv Mater 16 (2015) 043502. https://doi.org/10.1088/1468-6996/16/4/043502

[10] M. Oehme, E. Kasper, D. Weißhaupt, E. Sigle, T. Hersperger, M. Wanitzek, D. Schwarz, Two-dimensional hole gases in SiGeSn alloys, Semiconductor Science and Technology 37 (2022) 055009. https://doi.org/10.1088/1361-6641/ac61fe

[11] K. Han, Y. Wu, Y.C. Huang, S. Xu, A. Kumar, E. Kong, Y. Kang, J. Zhang, C. Wang, H. Xu, C. Sun, X. Gong, 2019, First Demonstration of Complementary FinFETs and Tunneling FinFETs Co-Integrated on a 200 mm GeSnOI Substrate: A Pathway towards Future Hybrid Nano-electronics Systems, 2019 Symposium on VLSI Technology, T182-T183. https://doi.org/10.23919/vlsit.2019.8776539

[12] D. Lei, K.H. Lee, S. Bao, W. Wang, B. Wang, X. Gong, C.S. Tan, Y.-C. Yeo, GeSn-on-insulator substrate formed by direct wafer bonding, Appl Phys Lett 109 (2016) 022106. http://dx.doi.org/10.1063/1.4958844

[13] T. Maeda, W.H. Chang, T. Irisawa, H. Ishii, H. Oka, M. Kurosawa, Y. Imai, O. Nakatsuka, N. Uchida, Ultra-thin germanium-tin on insulator structure through direct bonding technique, Semiconductor Science and Technology 33 (2018) 124002. https://doi.org/10.1088/1361-6641/aae620

[14] G. Lin, P. Cui, T. Wang, R. Hickey, J. Zhang, H. Zhao, J. Kolodzey, Y. Zeng, Fabrication of Germanium Tin Microstructures Through Inductively Coupled Plasma Dry Etching, IEEE Transactions on Nanotechnology 20 (2021) 846-851. https://doi.org/10.1109/TNANO.2021.3115509

[15] M. Wanitzek, M. Oehme, C. Spieth, D. Schwarz, L. Seidel, J. Schulze, 2022, GeSn-on-Si Avalanche Photodiodes for Short-Wave Infrared Detection, ESSCIRC 2022- IEEE 48th European Solid State Circuits Conference (ESSCIRC), 169-172. https://doi.org/10.1109/esscirc55480.2022.9911363

[16] R. Roucka, J. Tolle, C. Cook, A.V.G. Chizmeshya, J. Kouvetakis, V. D'Costa, J. Menendez, Z.D. Chen, S. Zollner, Versatile buffer layer architectures based on Ge1−xSnx alloys, Appl Phys Lett 86 (2005) 191912. https://doi.org/10.1063/1.1922078

[17] M. Bauer, J. Taraci, J. Tolle, A.V.G. Chizmeshya, S. Zollner, D.J. Smith, J. Menendez, C. Hu, J. Kouvetakis, Ge–Sn semiconductors for band-gap and lattice engineering, Appl Phys Lett 81 (2002) 2992-2994. https://doi.org/10.1063/1.1515133

[18] D.D.M. Wayner;, R.A. Wolkow, Organic modification of hydrogen terminated silicon surfaces, Journal of the Chemical Society, Perkin Transactions 2 (2002) 23-34. https://doi.org/10.1039/B100704L

[19] Max-Planck-Institut für Plasmaphysik, M. Mayer, SIMNRA, 2019, https://mam.home.ipp.mpg.de/

[20] E. Kasper, J. Werner, M. Oehme, S. Escoubas, N. Burle, J. Schulze, Growth of silicon based germanium tin alloys, Thin Solid Films 520 (2012) 3195-3200. https://doi.org/10.1016/j.tsf.2011.10.114

[21] H. Ye, J. Yu, Germanium epitaxy on silicon, Sci Technol Adv Mater 15 (2014) 024601. https://doi.org/10.1088/1468-6996/15/2/024601

[22] O. Steuer, D. Schwarz, M. Oehme, J. Schulze, H. Maczko, R. Kudrawiec, I.A. Fischer, R. Heller, R. Hubner, M.M. Khan, Y.M. Georgiev, S. Zhou, M. Helm, S. Prucnal, Band-gap and strain engineering in GeSn alloys using post-growth pulsed laser melting, J Phys Condens Matter 35 (2022) 055302. https://doi.org/10.1088/1361-648X/aca3ea

[23] O. Steuer, M.O. Liedke, M. Butterling, D. Schwarz, J. Schulze, Z. Li, A. Wagner, I.A. Fischer, R. Hubner, S. Zhou, M. Helm, G. Cuniberti, Y.M. Georgiev, S. Prucnal, Evolution of point defects in pulsed-laser-melted Ge(1-x)Sn(x)probed by positron annihilation lifetime spectroscopy, J Phys Condens Matter 36 (2023) 085701. https://doi.org/10.1088/1361-648X/ad0a10

[24] S. Abdi, S. Assali, M.R.M. Atalla, S. Koelling, J.M. Warrender, O. Moutanabbir, Recrystallization and interdiffusion processes in laser-annealed strain-relaxed metastable Ge0.89Sn0.11, Journal of Applied Physics 131 (2022) 105304. https://doi.org/10.1063/5.0077331

[25] P. Aella, C. Cook, J. Tolle, S. Zollner, A.V.G. Chizmeshya, J. Kouvetakis, Optical and structural properties of SixSnyGe1−x−y alloys, Appl Phys Lett 84 (2004) 888-890. https://doi.org/10.1063/1.1645324

[26] D.D. Cannon, J. Liu, Y. Ishikawa, K. Wada, D.T. Danielson, S. Jongthammanurak, J. Michel, L.C. Kimerling, Tensile strained epitaxial Ge films on Si(100) substrates with potential application inL-band telecommunications, Appl Phys Lett 84 (2004) 906-908. https://doi.org/10.1063/1.1645677

[27] R.R. Lieten, J.W. Seo, S. Decoster, A. Vantomme, S. Peters, K.C. Bustillo, E.E. Haller, M. Menghini, J.P. Locquet, Tensile strained GeSn on Si by solid phase epitaxy, Appl Phys Lett 102 (2013) 052106. https://doi.org/10.1063/1.4790302